\documentclass[
showpacs,
floatfix,
aps,
prl,
amsmath,
twocolumn,
superscriptaddress,
tightenlines
groupaddress,
]{revtex4-1}
\usepackage{amsmath,amssymb}
\usepackage{amscd, latexsym}
\usepackage{mathrsfs}
\usepackage{graphicx} 
\usepackage{epstopdf} 
\usepackage{cancel} 
\usepackage{amsfonts}
\usepackage{fancyhdr}
\usepackage{exscale}                
\usepackage{dcolumn}
\usepackage{bm}

\newcommand{\req}[1]{Eq.~(\ref{#1})}
\newcommand{\reqs}[1]{Eqs.~(\ref{#1})}

\newcommand{\be}{\begin{equation}}
\newcommand{\ee}{\end{equation}}
\newcommand{\bea}{\begin{eqnarray}}
\newcommand{\eea}{\end{eqnarray}}

\begin{document} 

\title{Effect of Ohmic environment on optimally controlled flux-biased phase qubit}

\author{Amrit Poudel}
\email{poudel@wisc.edu}
\author{Maxim G. Vavilov}
\email{vavilov@physics.wisc.edu}
\affiliation{Department of Physics, University of Wisconsin-Madison, Wisconsin 53706, USA }

\date{\today}

\begin{abstract}
We analyze the effect of environment on the gate operation of flux-biased phase qubits.  We employ the master  equation for a reduced density matrix of the qubit system coupled to an Ohmic environment, described by the Caldeira-Leggett model. 
Numerically solving this equation, we evaluate the gate error as a function of energy splitting between qubit states, junction capacitance, and temperature. 
The analysis is presented for single-quadrature microwave (control) pulses as well as for two-quadrature pulses, 
which lower the gate error significantly for idealized systems in the absence of environment. Our results indicate that two-quadrature pulses outperform single quadrature pulses even in the presence of environment.
\end{abstract}

\maketitle

Superconducting circuits containing Josephson junctions are promising candidates for scalable quantum information processing~\cite{Vion02,Devoret04,Clarke08}. However, small separations between successive quantum energy states in these circuits~\cite{Martinis02, Chiorescu03} do not permit selective manipulation of the qubit in a two dimensional subspace and results in a dynamical leakage of a quantum state to a broader Hilbert space  of the circuit~\cite{Steffen03}. To reduce this leakage, Motzoi {\it et al.} Ref.~\cite{Motzoi09} proposed a Derivative Removal by Adiabatic Gate (DRAG) method, which reduces the gate error to $10^{-5}$ for an experimentally optimal gate time of $6$\,ns. This error  is well below the required error threshold of $10^{-3}$ for fault tolerant quantum computation~\cite{Knill05}.

In addition to the dynamic leakage, any realistic model of a qubit must also address coupling of the qubit to environment, which leads to further destruction of qubit states. Several efforts have already been made towards the study of accurate control of a qubit system~\cite{Steffen03,Rebentrost09, 2Rebentrost09}. However, the effect of an environment on optimally controlled qubit has only been studied in a phenomological model~\cite{Motzoi09}, which leads to the evolution of density matrix of the qubit in Lindblad form~\cite{Lindblad76}. 

In this paper, we resort to a microscopic approach to the modeling of the environment. We employ the Caldeira-Leggett model of the system-environment coupling~\cite{Caldeira83,Leggett87} to describe time evolution of a flux-biased phase qubit, driven by the DRAG pulses~\cite{Motzoi09}. Numerically solving equation of motion for the qubit density matrix, we study the dependence of the gate error on temperature and environmental coupling strength. 

\emph{Model}.
A flux-biased phase qubit consists of a Josephson junction (JJ) embedded in a superconducting loop~\cite{Devoret04}. Finite resistance of the JJ results in dissipation processes in the qubit and can be accounted for by the Caldeira-Leggett model~\cite{Caldeira83,Leggett87}. The full Hamiltonian of the qubit and the environment is 
\be
\hat H = \hat  H_q+ \hat  P(t) + \hat  H_R +\hat  V.
\ee 
The Hamiltonian of the qubit $\hat H_q$ is written in terms of operators $\hat Q$ and $\hat \delta$, the charge and phase difference of the JJ respectively:
\begin{equation}
    \label{eq:tid-H}
    \hat H_q = \frac{\hat  Q^2}{2C} + \frac{\phi_0}{2\pi}\left[\frac{\phi_0}{4\pi L}\left(\hat  \delta - \frac{2\pi\phi_{ext}}{\phi_0} \right)^2 - I_0\cos\hat  \delta\right]\;,
\end{equation}
where $L$ ($C$) is the loop inductance (junction capacitance),
$\phi_{ext}$ is the external magnetic flux applied to the phase qubit, 
$I_0$  is the critical current of the JJ, and $\phi_0 = h/2e$ is the flux quantum. The qubit is capacitively coupled to microwave current source, used to induce coherent transitions between the qubit states~\cite{Devoret04}. This coupling introduces time-dependent part in the Hamiltonian: 
\begin{align}
    \label{eq:td-H}
\hat P(t) &= \frac{\phi_0 I(t)}{2\pi}  \hat{\delta}\;.
\end{align}
Here $I(t)= I_x(t)\cos \omega_d t + I_y(t)\sin \omega_d t$ is microwave current with frequency $\omega_d$. 

The environment is introduced as a set of harmonic oscillators (reservoir) with the Hamiltonian 
$\hat H_R =  \sum_{\alpha=1}^N (m_\alpha/2)\left(\hat p_\alpha^2/m_\alpha^2 + \omega_\alpha^2 \hat x_\alpha^2\right)$. The coupling between the qubit system and the reservoir is bilinear in the JJ phase $\hat{\delta}$  and oscillator displacements $\hat x_\alpha$: 
\be 
 \hat V = \sum \limits_{\alpha=1}^N \gamma_\alpha \hat x_\alpha \hat q \;, \quad
 \hat q \equiv \hat \delta-\frac{2\pi\phi_{ext}}{\phi_0}\; ,
\ee
where parameters $\gamma_{\alpha}$ determine the coupling strength between the qubit and reservoir mode $\alpha$.

Our goal is to describe the time evolution of the qubit density matrix $\hat \rho(t)$. The qubit is initially prepared in a pure state, corresponding to the density matrix $\hat \rho(0)$. Assuming that the environment is in a thermal equilibrium at temperature $T$, the master equation for $\hat \rho(t)$ has the following form~\cite{Paz01}: 
\begin{align}
    \label{eq:masterEq}
    \frac{d\hat \rho}{dt} = \frac{1}{i\hbar} \Big[\hat H_q(t),\, \hat \rho(t)\Big] - 
    \hat{\mathcal{L}}_t \{  \hat \rho\} \;,   
\end{align}
and the dissipative term is
\begin{align}
    \hat{\mathcal{L}} _t\{\hat \rho\} \equiv \frac{1}{\hbar^2}\int_{0}^{t}\,dt'\,\eta_1(t')\Big[\hat q\,,\big[\hat{\tilde{q}}(-t'),\,\hat \rho\big]\Big]\, \nonumber \\
      - \frac{1}{\hbar^2}\int_{0}^{t}\,dt'\,\eta_2(t')\Big[\hat q\,,\{\hat{\tilde{q}}(-t'),\,\hat \rho\}\Big] \; , 
\end{align}
where $\hat{\tilde{q}}(t)$ is a Heisenberg operator. In~\req{eq:masterEq}, $\eta_1(t)$ accounts for dissipative part of the dynamics and $\eta_2(t)$ represents the quantum noise of the environment~\cite{WeissBook}:
\begin{align}
    \eta_1(t) &= \hbar \int_0^{\infty} J(\omega)\left[1 + 2\,N(\omega)\right] \cos \omega t \, d\omega\;, \\
    \eta_2(t) &= i\hbar \int_0^{\infty} J(\omega) \sin \omega t\, d\omega \;.
\end{align}
The spectral density $J(\omega) = \sum_{\alpha=1}^N\gamma_\alpha^2/(2m_\alpha \omega_\alpha)\delta(\omega-\omega_\alpha)$ for an ohmic environment is
\begin{equation}
J(\omega) = \frac{C\hbar^2}{4e^2} \omega_0 \xi \omega e^{-\omega/\omega_s}\;, 
\end{equation}
where $\xi$ is a dimensionless parameter, and $\omega_s$ is a cutoff frequency that exceeds 
all other frequency scales of the system. The Planck's function
$N(\omega) = 1/[\exp(\hbar\omega/T) -1]$ defines an average excitation number of environment modes with frequency $\omega$. 

In experimental setup~\cite{McDermott10}, the ''potential'' part of $\hat H_q$ in~\req{eq:tid-H} has one deep minimum and another very shallow minimum that disappears at the critical flux $\phi_c$. 
External flux $\phi_{\rm ext}$ is chosen in  such a way that only a few  levels are localized in the shallow well, but these levels are still separated from levels localized in the deep well by impenetrable barrier~\footnote{The levels in the deep well can also be accounted in the present model, however, our numerical results indicate that the gate error does not change significantly if those levels are also included in the calculation for chosen values of parameters.}. As a result, we truncate the qubit Hamiltonian,~\reqs{eq:tid-H} and~\eqref{eq:td-H}, to three localized levels and obtain the following Hamiltonian in energy-representation:
\begin{align}
\label{eq:Hamt}
H_q(t) = \hbar \sum_{j =1}^2 \left[\omega_{j-1}\hat \Pi_j + a \lambda_j \hat \sigma_j^+ + a^{*} \lambda_j 
\hat \sigma_j^- \right] + \hat H_{nr} \;,
\end{align}
where $\hat \Pi_j = |j\rangle\langle j|$ is the projector for the $j^{th}$ level, $\hat \sigma_j^+ = |j\rangle\langle j-1|$ is the raising operator,  $a = (I_x - iI_y)e^{i\omega_dt}/2$ is the amplitude of microwave drive, $\lambda_j=\phi_0 \langle j| \hat \delta|j-1\rangle/2\pi\hbar$ is the matrix element of the phase operator, 
$\omega_j = (\varepsilon_{j+1}-\varepsilon_0)/\hbar$, $\varepsilon_j$ is an energy eigenvalue of time-independent Hamiltonian $H_q$ and $\hat H_{nr}$ contains non-resonant terms. In this three-level model, the lower two energy levels comprise qubit space while the third level accounts for a leakage level.
 
\emph{Gate error and DRAG method}.
In order to quantify the error during gate operation we use gate fidelity averaged over two initial input states in a two dimensional Hilbert space, similar to one defined in Ref.~\cite{Bowdrey02}:
\begin{equation}
F_g = \frac{1}{2} \sum_{j=1}^2 Tr\left[\hat U_{ideal}\hat \rho^{(0)}_j \hat U_{ideal} ^{\dag} \hat \rho_j(t_g)\right] \;.
\end{equation}
Here $\hat U_{ideal}$ represents an ideal evolution, $\hat \rho_j(t)$ is an actual density matrix of a qubit system with $\hat \rho_j(0)=\hat\rho^{(0)}_j$,
and $\hat\rho^{(0)}_j$ represents two initial axial states in a Bloch sphere. The gate error $E$ is defined as $E = 1 - F_g$. 

A simple approach to minimize leakage of quantum information from qubit subspace is to use a single-quadrature Gaussian envelope pulse given by 
\begin{align}
I_x(t) = I_{\pi}(t) = A\,e^{-\left(t-t_g/2\right)^2/2\sigma^2} - B\;,\;\; I_y(t) = 0 \;,
\label{eq:gauss1}	
\end{align}
where $t_g$ is a gate time and $\sigma = t_g/2$.  
For a NOT gate operation, which we choose to focus on without any loss of generality, constant $A$ is defined by $\int_0^{t_g} I_{\pi}(t)\,dt = \pi$ and $B$ is chosen so that the Gaussian pulse starts and finishes off at zero. However, such a pulse shape still suffers a large gate error of about $10^{-2} $  as shown below. 

The DRAG method reduces the gate error to order of $10^{-5}$ for a gate time of $6$\,ns~\cite{Motzoi09} by  using two quadratures and time-dependent detuning $d_1(t)=\omega_0-\omega_d=(\lambda^2-4)I_\pi^2(t)/4\Delta$, where the anharmonicity parameter $\Delta \equiv \omega_1 - 2\,\omega_0$, and $\lambda$ measures relative strength of $0 \to 1$ and $1 \to 2$ transitions, that is, $\lambda \equiv \lambda_2/\lambda_1$. We note that the laboratory frame is more suitable for the solution of the reduced density matrix of the qubit coupled to environment. We preserve the form of the quadrature amplitudes as in Ref.~\cite{Motzoi09} for a Hamiltonian in rotating frame
\begin{subequations}
\begin{align}
	I_x  &= I_{\pi} + \frac{(\lambda^2 -4) I^3_{\pi} } {8 \Delta^2} \;, \ \ I_y  = \frac{-\dot{I}_{\pi} } {\Delta}\;.
	 \label{eq:DRAG}
\end{align}
Then,  we obtain the following equation for the microwave driving frequency for the Hamiltonian  \req{eq:td-H} in the laboratory frame 
\begin{align}
	 t \dot \omega_d(t)+ \omega_d(t)  = \omega_0-d_1(t)\;,  \quad \omega_d(0) = \omega_0\;. 
\end{align}
\end{subequations}

\begin{figure} [t]
\includegraphics[width=0.48\columnwidth]{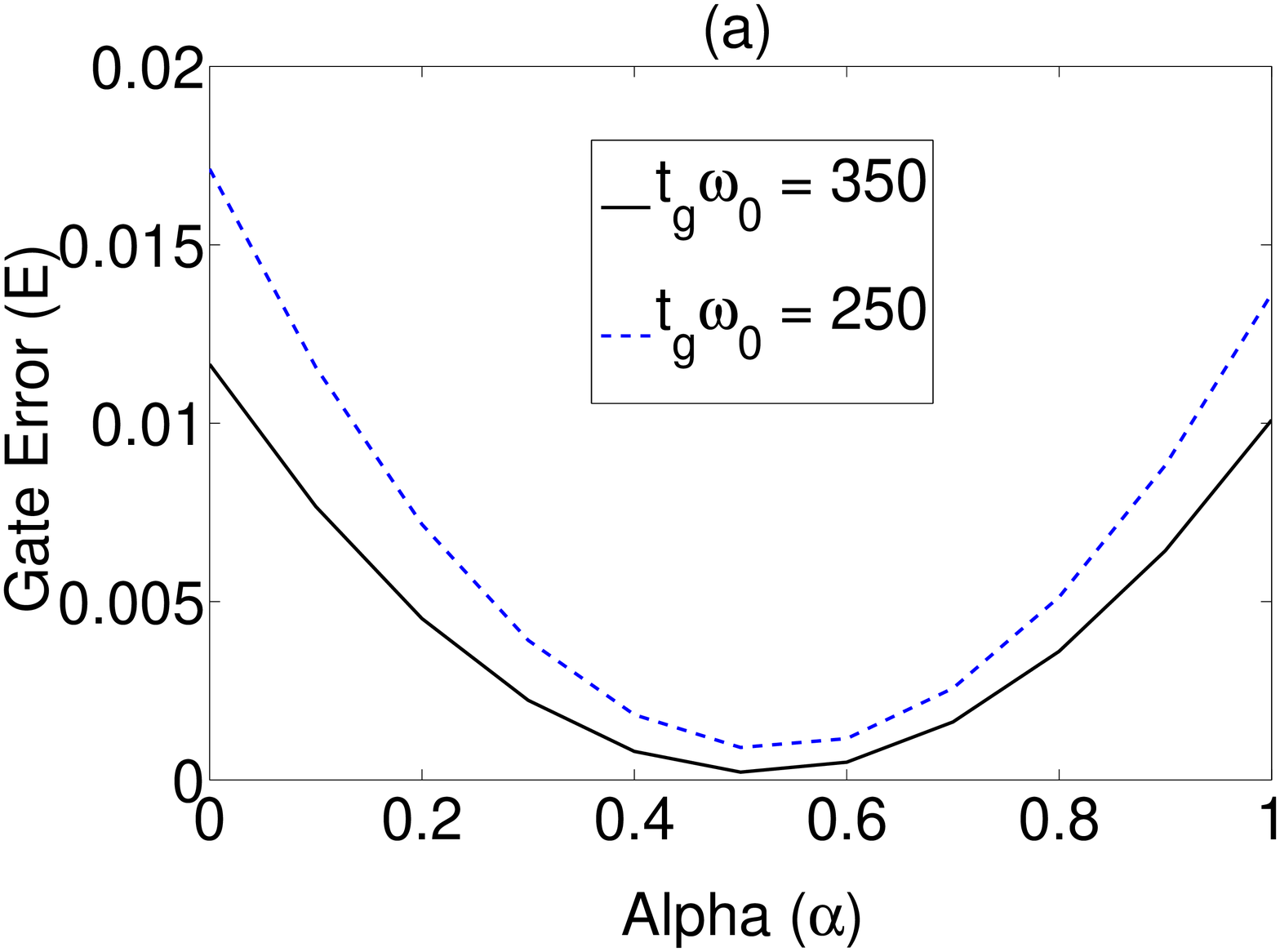} 
\includegraphics[width=0.48\columnwidth]{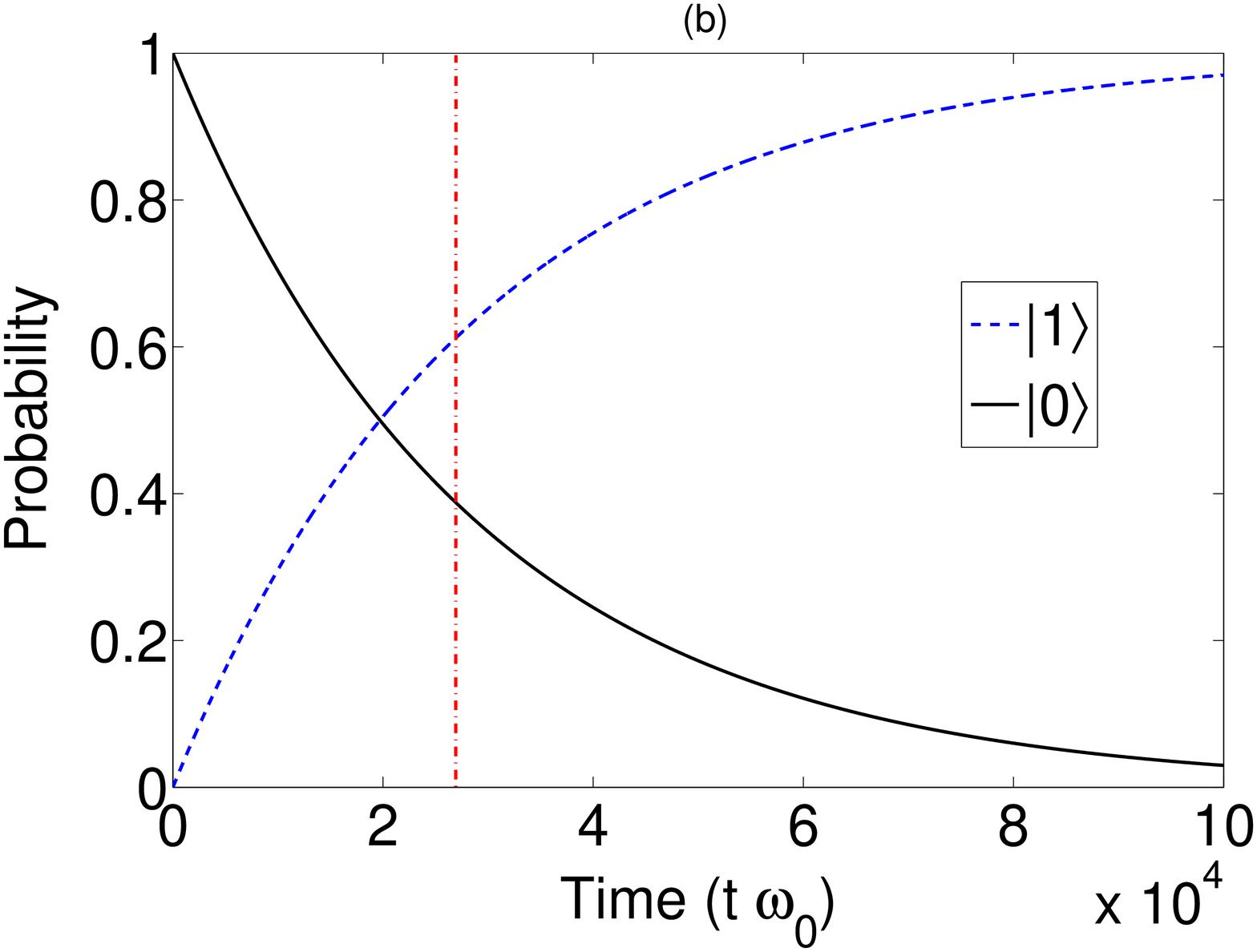}
\caption{(Colors online) (a) Gate Error vs.~Alpha for Gate Times $t_g\omega_0 = 250$ (dashed blue) and $t_g\omega_0 = 350$ (solid black). (b) Probability vs.~Time for Temperature  $T = 0.1\,\hbar \omega_0$ and $\xi = 2$. The qubit is initially prepared in $|1\rangle$ state (solid black), which relaxes to $|0\rangle$ state (dashed blue) due to dissipation. }
\label{f1}
\end{figure}

Although the DRAG correction is successful in reducing the gate error below the required threshold, a practical implementation may not be feasible  due to stringent requirement to vary microwave frequency. For this reason, we also consider two-quadrature pulses  with fixed driving frequency $\omega_d = \omega_0$~\footnote{We also made constant detuning of the driving frequency from $\omega_0$, but did not see any improvement compared to $\omega_d = \omega_0$ case.}. To obtain relation between $I_{x,y}(t)$ components, we perform an adiabatic transformation 
$\hat D = \exp\left[-iI_x \alpha/2\Delta \left(\hat\sigma^y_{1} + \lambda \hat \sigma^y_{2} \right) \right]$
of the rotating frame Hamiltonian as in Ref.~\cite{Motzoi09} along with a dimensionless parameter $\alpha$, and $\hat \sigma^y_i=-i[\hat{\sigma}^+_i-\hat{\sigma}^-_i]$. We then require that the imaginary parts of $0\to1$ and $1\to2$ matrix elements of the transformed Hamiltonian vanish, while $1\to1$ matrix element is reduced to the second order in $I_x(t)/\Delta$ and obtain:
\begin{align}
	I_y = \frac{-\alpha\dot{I}_\pi } {\Delta} \;,  \quad
	\alpha = \frac{\lambda^2}{4}   \;. 
	\label{eq:dragfixed}
\end{align}
For the phase qubit, $\lambda \approx \sqrt{2}$, which implies $\alpha = 0.5$ (see Ref.~\cite{Chow10} for transmon qubits for which $\alpha = 0.4$ owing to different value of $\lambda$). Similar result is obtained from direct numerical simulation of the gate error for different values of $\alpha$. As shown in Fig.~\ref{f1}(a), a minimum value of the error occurs at around $\alpha = 0.5$ for gate times $t_g\omega_0 = 250$ (dashed blue) and $t_g\omega_0 = 350$ (solid black). 

\emph{Results}. 
Numerical parameters used below in our simulation are chosen accordingly to the actual experimental setup: 
$C = 1$\,pF, $I_0 = 1.5\,\mu$\,A, $\beta_L = 2\pi I_0 L/\phi_0 = 3.2$, and $\phi_{ext} = 0.955\,\phi_c$\,, where $\phi_c$ is a critical flux~\cite{McDermott10}. For these parameters, we find $\omega_0 \approx 39$\,GHz, $\lambda \approx 1.41$ and $\Delta   \approx -2.4$\,GHz.   

\begin{figure} [t]
\includegraphics[width=0.9\columnwidth]{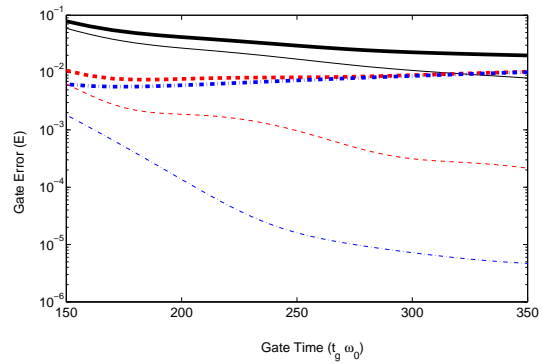}
\caption{(Colors online) Gate Error vs.~Gate Time with (thick lines) and without (thin lines) dissipation for Gaussian ($\sigma = 0.5\,t_g$) (solid black), Gaussian ($\sigma= 0.5\, t_g$) with first order DRAG correction and time-dependent driving frequency (dashed-dot blue), and Gaussian ($\sigma = 0.5\,t_g$) with $\omega_d = \omega_0$ and $\alpha = 0.5$ (dashed red), all in the laboratory frame.}
\label{f2}
\end{figure}

In Fig.~\ref{f2}, we plotted the gate error for the DRAG pulses with and without time-dependent detuning. We find that pulses with two quadratures and fixed driving frequency (thin dashed red) perform much better than single quadrature Gaussian pulses (thin solid black), but are not as effective as pulses with double quadratures and time-dependent driving frequency (thin dashed-dot blue). 

In order to study the effect of dissipation on the DRAG pulses, we solve the master equation~\eqref{eq:masterEq} numerically. First, we consider the relaxation of the qubit from the first excited state to the ground state in the absence of microwave drive, which is shown in Fig.~\ref{f1}(b). For this simulation, we choose parameter $\xi = 2$ so that the relaxation time $T_1 \approx 700$\,ns  corresponds to experimentally observed decay time for phase qubits \cite{Martinis09}. We note that the spontaneous relaxation rate of the first excited state can also be evaluated from the master equation~\eqref{eq:masterEq}
\begin{equation}
 \Gamma = \frac{1}{T_1}=2\pi\hbar  \omega_0^2 \frac{\xi C}{4e^2}|q_{01}|^2, \quad 
 q_{01}=\langle 0|\hat q|1\rangle.
 \end{equation}

For the above choice of dimensionless coupling parameter $\xi$ we study the effect of dissipation on the DRAG corrections. In Fig.~\ref{f2}, we observe a non-monotonic behavior of the gate error with gate time for pulses with the DRAG corrections.  We find that for shorter gate times, the DRAG correction with time-dependent driving frequency is less affected by dissipation (thick dashed-dot blue). However, for longer gate times, dissipation has a substantial effect on two-quadrature pulses. For $t_g\omega_0 = 250$ ($t_g \approx 6$ ns), the gate error increases from  $10^{-5}$  to higher order of $10^{-3}$ when dissipation is turned on for the same DRAG pulses with dynamical detuning. This increase in the gate error is due to the relaxation of the qubit from the excited state to the ground state, which becomes prominent for longer gate times.  For comparison, we plotted the gate error for three different types of pulses: single quadrature Gaussian pulse (thick solid black), the Gaussian pulse with first order DRAG correction and time-dependent driving frequency (thick dashed-dot blue) and the Gaussian pulse with the DRAG correction and resonant driving frequency (thick dashed red). One can conclude from these plots that the performance of two-quadrature pulses without detuning is comparable to the DRAG pulses with dynamical detuning when dissipation is turned on.

\begin{figure} [t]
\includegraphics[width=0.9\columnwidth]{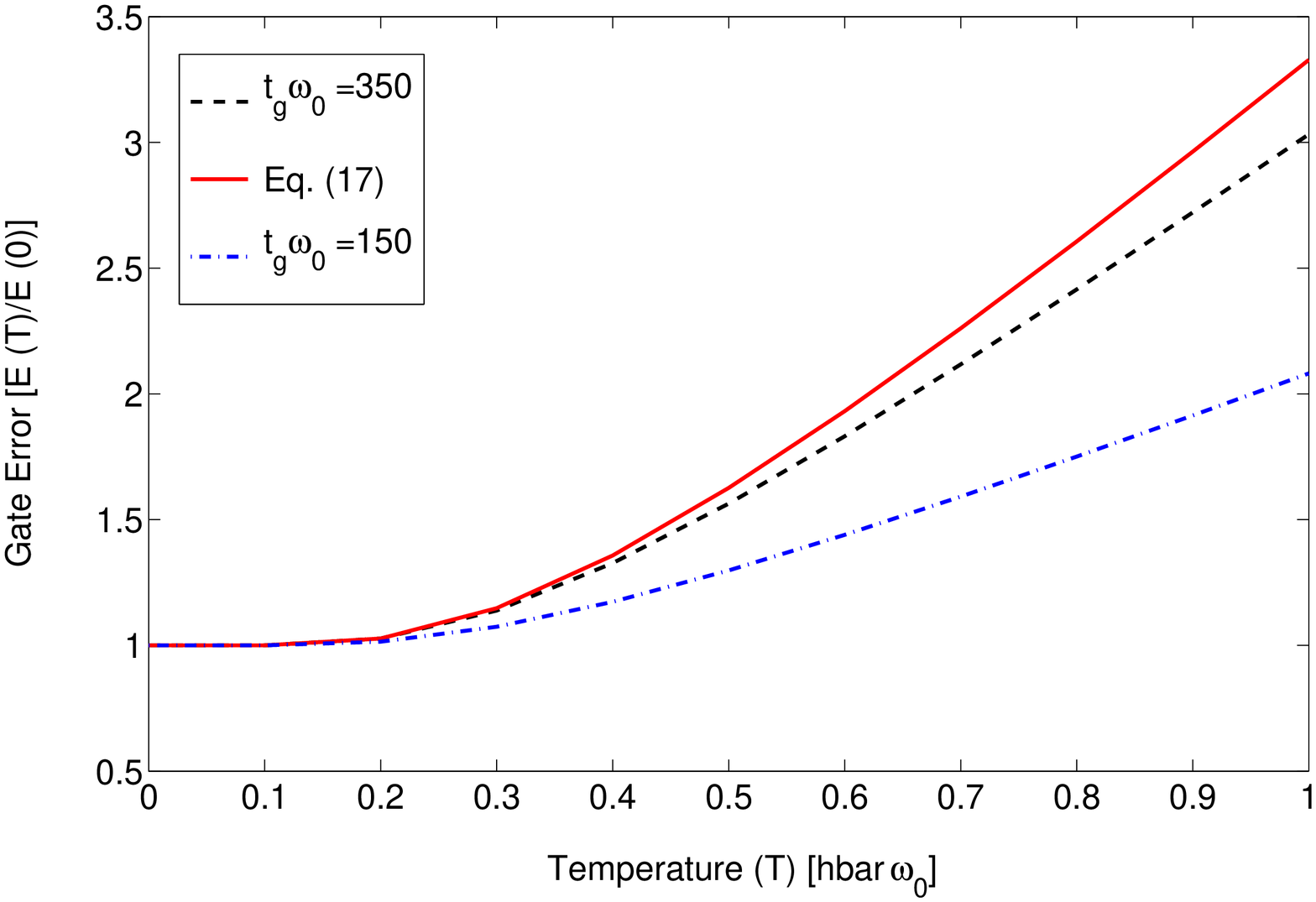} 
\caption{(Colors online) Normalized Gate Error vs.~Temperature for Gaussian ($\sigma = 0.5\, t_g$) with the DRAG correction and time-dependent driving frequency from the numerical simulation for gate times $t_g\omega_0 = 150$ (dashed-dot blue) and $t_g\omega_0 = 350$ (dashed black). Analytical rate equation estimation of the normalized gate error (solid red). }
\label{f3}
\end{figure}

Next we study the effect of temperature on the gate error. The plot (Fig.~\ref{f3}) for the gate error normalized around error at zero temperature reveals a monotonic increase of the gate error with temperature. The monotonic increase in the gate error is due to increase in the relaxation rate with temperature. In Fig.~\ref{f3}, we plotted the gate error for two different gate times: $t_g\omega_0  = 150$ (dashed-dot blue) and $t_g\omega_0 = 350$ (dashed black). 

We now compare the results of numerical solution of the master equation~\eqref{eq:masterEq}, and the simple picture of the error due to coupling to the environment in terms of the ''Fermi-Golden rule'' transition rates. Considering an environment at zero temperature and assuming that the contribution  to the error $E$ from the environment is small, $E\ll 1$, we can evaluate the error as the probability of an excitation of a reservoir mode
during the  qubit operation, which happens with rate $\Gamma$: $E(T=0)= \Gamma t_g \overline{\rho_{11}(t)}$, where $\overline{\rho_{11}(t)}=\int_0^{t_g}\rho_{11}(t)dt/t_g$ is the time-average of probability of qubit being in the first excited state. At finite temperature, the processes with excitation of environment happen with rate $\Gamma(T)=\Gamma[1+N(\omega)]$. In addition, the qubit can absorb an excitation from the environment with rate $\Gamma N(\omega)$.  Combining these processes, we obtain the following estimate for the gate error due to coupling to the environment: 
\be
\begin{split}
\frac{E(T)}{ \Gamma t_g} = & 
\left[\{1+N(\omega_0)\}+\lambda^2N(\omega_1-\omega_0)\right] \overline{\rho_{11}(t)}\\
&+N(\omega_0) \overline{\rho_{00}(t)} \;.
\end{split}
\ee
For average occupation of the ground and the first excited states being $\approx 1/2$, and for a weak anharmonicity of the qubit system $|\Delta|\ll \omega_0$, the gate error reduces to
\be
\frac{E(T)} {E(0)} \approx  1 + 4 N(\omega_0)\;.
\ee
The estimated normalized gate error (solid red) is plotted in Fig.~\ref{f3} together with the gate error obtained from numerical simulation. The rate equation estimation of the error is fairly close to the error obtained from direct numerical simulation for a longer gate time (dashed black). However, for a shorter gate time (dashed-dot blue), the estimated error deviates from the exact numerical simulation considerably suggesting that the rate equation description may not be valid for shorter gate times and higher temperatures.

\emph{Discussion and Conclusions}. 
In comparison to single-quadrature pulses, two-quadrature microwave (control) pulses lead to significant suppression of the gate error. Despite the presence of dissipative environment, two-quadrature pulses reduce the gate error close to the desired threshold for fault tolerant quantum computation. At the same time, we also determined that optimal two-quadrature pulses with fixed microwave frequency provide similar level of the gate error as two-quadrature pulses with time-varying frequency.  

In addition, we observed a monotonic increase of the gate error with temperature, which is due to increase in the relaxation rate with temperature. We found that temperature dependence of the gate error for longer pulses can be captured by a simple error estimation based on the rate equations. Nonetheless, the simple estimation of the error for shorter pulses differs from the gate error obtained from direct numerical solution of the reduced density matrix. Hence, we concluded that full density matrix solution is necessary to calculate the error for shorter gate times. 

\emph{Acknowledgements.}
We are grateful to Robert Joynt and Robert McDermott for fruitful discussions. The work was supported by NSF Grant No. DMR-0955500.

\end{document}